\documentclass[11pt]{article}
\usepackage[margin=1in]{geometry}
\usepackage{graphicx}

\def\beq{\begin{equation}}
\def\eeq#1{\label{#1}\end{equation}}
\def\eeqn{\end{equation}}

\def\beqa{\begin{eqnarray}}
\def\eeqa#1{\label{#1}\end{eqnarray}}
\def\eeqan{\end{eqnarray}}

\let\bar=\overbar

\def\Dslash{\not{\hbox{\kern-4pt $D$}}}
\def\dslash{\not{\hbox{\kern-2pt $\del$}}}

\def\msb{{\bar{\ssstyle M \kern -1pt S}}}

\def\Title#1{\begin{center} {\Large {\bf #1} } \end{center}}
\def\Author#1{\begin{center} {\normalsize {\sc #1} } \end{center}}
\def\Institution#1{\begin{center} {\normalsize {\it #1} } \end{center}}
\def\Abstract#1{\noindent {\normalsize {\bf Abstract:} {\normalfont #1}}}
\def\Conference{\vspace{4mm}\begin{raggedright} {\normalsize {\it Talk presented at the 2019 Meeting of the Division of Particles and Fields of the American Physical Society (DPF2019), July 29--August 2, 2019, Northeastern University, Boston, C1907293.} } \end{raggedright}\vspace{4mm}}

\begin{document}

%
%

\Title{Barium Tagging with Selective, Dry-Functional, Single Molecule Sensitive On-Off Fluorophores for the NEXT Experiment}

\Author{N.K. Byrnes\footnote{Presenter and Corresponding Author: byrnes.nicholas@mavs.uta.edu}, A. A. Denisenko, F.W. Foss Jr., B.J.P. Jones, A.D. McDonald, D.R. Nygren, P. Thapa, K. Woodruff for the NEXT Collaboration}

\Institution{Department of Physics, University of Texas at Arlington}

\Abstract{In the search for neutrinoless double beta decay, understanding and reducing backgrounds is crucial for success. An advance that could drive backgrounds to negligible levels would be the ability to efficiently detect the barium daughter in $^{136}$Xe to $^{136}$Ba double beta decay, since no conventional radioactive process can produce barium ions or atoms in xenon at significant rates. In xenon gas, the barium daughter most likely survives as a dication. An approach under development by the NEXT collaboration involves transporting this ion from the active medium onto a coated transparent plane supporting a barium-sensitive fluorescent dye, monitored via fluorescence microscopy. Upon exposure to a barium dication, the dye will begin fluorescing, which, when correlated with the detection of a double electron signal at the anode, would confirm double beta decay. Our results have shown that a single barium ion can be resolved via Single Molecule Fluorescent Imaging (SMFI). The next challenge is a realization of this technique within in a large volume of xenon gas. Significant advances have recently been made:  custom barium-tagging molecules that fluoresce strongly in the dry state when exposed to barium have been demonstrated, and devices constructed that can observe fluorescence via in-vacuum or in-gas Total Internal Reflection Fluorescence Microscopy. We present the status of this technique and the outlook for barium tagging with On-Off switchable fluorophores, including new results with a Ba$^{2+}$-selective dye that functions under our desired conditions in the visible region and with single ion sensitivity.}

\Conference

%
%

\section{Introduction}

The search for lepton-number violating nuclear processes is a topic of great experimental interest. If such processes are observed, it would imply that the neutrino has a Majorana nature~\cite{Schechter:1980gr} - that is, it is its own antiparticle.  One of the strongest candidate processes is neutrinoless double beta decay ($0\nu\beta\beta$), $^N_A X\rightarrow_{A-2}^N X+2e^-$~\cite{Vergados:2012xy,Bilenky:2012qi}. Observation of this decay show simultaneously that lepton number must not be conserved in nature; there exists new physics beyond the standard model allowing for the lightness of neutrino masses~\cite{Chang:1985en}; and that leptogenesis is a possible mechanism for creating matter-antimatter asymmetry in the early Universe~\cite{Fukugita:1986hr}. Of interest in this regard is the decay of $^{136}$Xe to $^{136}$Ba. The quest to detect $0\nu\beta\beta$ from $^{136}$Xe has led to the development of a host of detector technologies allowing for increased sensitivity and background reduction. This is due to the extremely long limits placed on the lifetime of $0\nu\beta\beta$, on which present experiments using xenon place the lower limit at 1.07$\times 10^{26}$ years~\cite{KamLAND-Zen:2016pfg}, with current theoretical models predicting that it could be far longer~\cite{Vergados:2012xy,Bilenky:2012qi}. Detection of such a rare process is a difficult prospect, requiring precise detectors with sufficient shielding from cosmic radiation, excellent radio-purity, and extremely signal-selective identification techniques. One possible technique to reduce the background to negligible levels would be the ability to detect the daughter ion produced from $0\nu\beta\beta$~\cite{Moe:1991ik} in coincidence with energy resolution better than 2\% FWHM~\cite{Renner:2019pfe} to reject the two-neutrino mode background, possibly further enhanced with topological identification techniques~\cite{Ferrario:2019kwg}.  High pressure xenon gas appears to be an especially promising medium for satisfying these diverse requirements~\cite{Nygren:2009zz}.

\begin{figure}
\begin{centering}
\includegraphics[width=0.49\columnwidth]{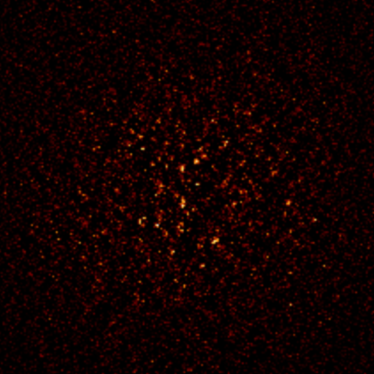}
\includegraphics[width=0.49\columnwidth]{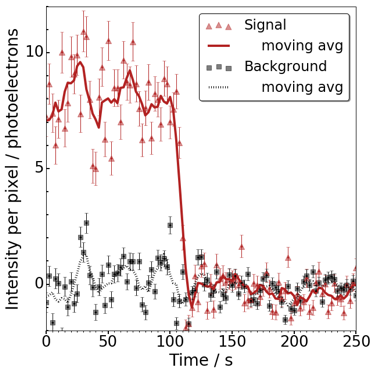}
\par\end{centering}
\caption{Image showing the fluorescent single molecules of FLUO-3 exposed to barium ions in a PVA matrix, treated with BAPTA (left) and fluorescence trajectory of a 5x5 pixel region of the EMCCD camera showing the single step photobleaching process of FLUO-3(right), from Ref.~\cite{McDonald:2017izm}.
\label{fig:SingleBaFig}}
\end{figure}

 The NEXT collaboration~\cite{Martin-Albo:2015rhw} is currently developing single molecule fluorescence imaging (SMFI)~\cite{stuurman2006imaging} techniques to efficiently tag the barium daughter dication from a $^{136}$Xe double beta decay as a way to suppress radiogenic backgrounds to vanishing levels.  This method was first outlined in Ref~\cite{Elba} and further developed in Ref~\cite{Jones:2016qiq}. Through the use of SMFI, it has been demonstrated that imaging of individual barium ions is possible using commercially available chemosensors in the FLUO family~\cite{McDonald:2017izm}, which are comprised of a carboxlylic acid based barium binding group connected to an on-off switchable fluorophore, fluorescein, via a nitrogen ''switch.''  Upon binding barium, the molecule transforms from a non-fluorescent state into a fluorescent one, and can be resolved at the single molecule level. Fig.~\ref{fig:SingleBaFig} shows individual barium ions resolved spatially with 2~nm super-resolution (left), and identified as single molecule emitters via their discrete photo-bleaching processes (right).  

Due to the nature of the final detection environment, the  fluorophores and microscope configurations employed in Ref.~\cite{McDonald:2017izm} must be augmented in order to realize a barium tagging system operable in xenon gas. These augmentations have been the subject of an interdisciplinary research effort between the physics and chemistry departments at the University of Texas Arlington. Particular areas of focus have been the development of a molecule capable of on-off fluorescence upon chelation of barium in the dry state, a technique for transporting the barium ion to the chemosensors, and a microscope operable within a high pressure xenon gas vessel. This proceeding presents the current status of these efforts.

\section{Novel Fluorophores for Dry Applications}

The limitations of the FLUO dyes for dry barium sensing include that the receptor uses carboylic acids that must deprotonate to accept dications, and that the fluor undergoes  configurational changes between ground and excited states that are suppressed in dry conditions.  In Ref~\cite{Thapa:2019zjk}, we proposed, synthesized and demonstrated custom-made crown-ether derivative fluorophores for dry barium ion sensing.  The receptor group 18-crown-6 has shown high selectivity towards barium in the past and does not require deprotonation in order to bind, suggesting promise for dry function. As such, its azo-derivative was identified as a promising receptor for dry fluorophore design that could also regulate the photophysical state of a neighboring fluorophore.  The switchable fluorescent group coupled to this receptor must also respond dry, for which small, rigid aromatics as the ideal candidates.

Of the candidate crown-ether / dye combinations explored, 18c6-py (18-crown-6 bonded to pyrene) and 18c6-an (18-crown-6 bonded to anthracene) showed the strongest barium-sensitive response, which was demonstrably maintained in the dry phase and compatible with fluorescence microscopy techniques.  While the anthracene species had more amenable properties, both pyrene and anthracene require UV excitation. This introduces intrinsic challenges due to ultraviolet excitation of background fluorescence in glass and quartz, and generation and steering of sufficient laser output power at short wavelengths.  Furthermore, pyrene and anthracene are highly photo-stable, which removes the tool of photo-bleaching, a vital property for identification of single molecule candidates with our previously demonstrated approaches.  These properties presented sufficient hurdles for single ion detection that a subsequent phase of fluorophore development was instigated, culminating in the dry-compatible system with single ion resolution described below.
\begin{figure}
\begin{centering}
\includegraphics[width=0.99\columnwidth]{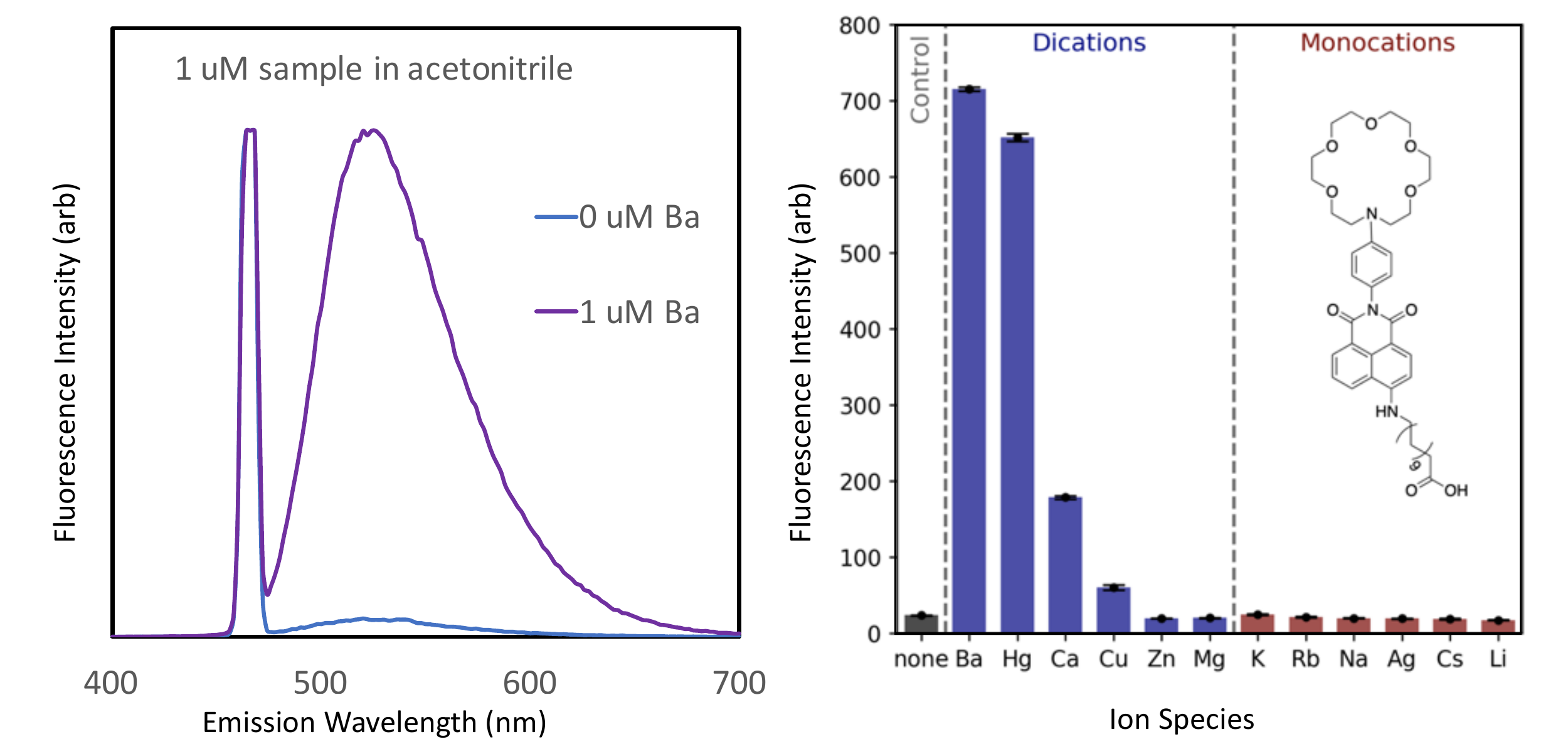}
\par\end{centering}
\caption{Fluorescent response of 18c6-nap to added Ba$^2{^+}$ and low constitutive fluorescence in the absence of additional Ba$^2{^+}$ (left); 1 $\mu$M 18c6-nap in the presence of 5 $\mu$M metal cations at room temperature demonstrating selective response for dications, particularly Ba$^2{^+}$ and Hg$^2{^+}$. (right}
\label{fig:Selectivity}
\end{figure}

Our goals in continued chemosensor development beyond Ref~\cite{Thapa:2019zjk} include moving from UV to visible excitation; continued improvement of the response difference between a barium sparse and barium enriched sample; re-introduction of photo-bleaching effects; and enhanced selectivity towards barium vs other ions.  Specifically, any residual monocation sensitivity would be very problematic for implementation in a gaseous $0\nu\beta\beta$ experiment due to possible chelation, even at a sub-leading level, with Xe$^+$.  To this end, we extended the 18c6 family of Ref~\cite{Thapa:2019zjk} to include a new molecule, 18c6-nap, built according to the same principles but using naphthalimide as the fluorescent group.  The structure of this molecule is shown in Fig.~\ref{fig:Selectivity}.

As shown in Fig.~\ref{fig:Selectivity}, 18c6-nap shows a strong barium response in the visible region between 500 and 620~nm and excitation peaking at 464~nm in non-aqueous solvent (acetonitrile). This large stokes shift is especially promising for fluorescence microscopy. Comparison of the response to other ionic species (Fig.~\ref{fig:Selectivity}, right), shows strong selectivity.  Unlike in the FLUO family, here barium out-competes calcium roughly four to one, and mercury is the only competitive species tested, likely due to its similar ionic size to barium.  Dications of other ions are very unlikely in coincidence with a double beta decay event, and no monocations produce a significant response suggesting that Xe$^+$ will not be a concern. This property represents an important advance in the drive toward molecules suitable for single barium ion detection in xenon gas.

\begin{figure}
\begin{centering}

\includegraphics[width=0.99\columnwidth]{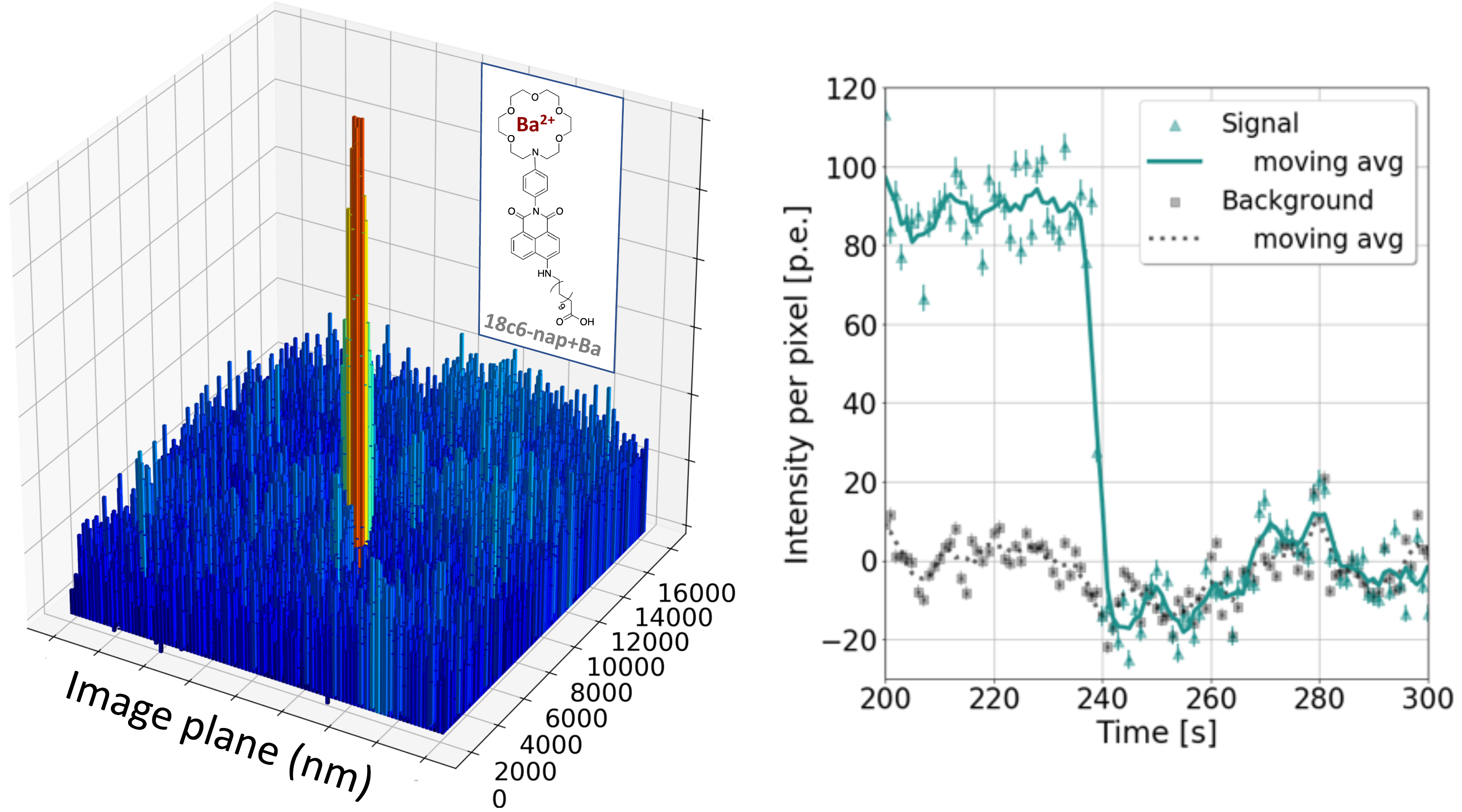}

\par\end{centering}
\caption{Image showing a single detected molecule of 18c6-nap exposed to barium ions in a PVA matrix (left), and fluorescence trajectory of a 5x5 pixel region of the EMCCD camera showing the single step photobleaching process of 18c6-nap (right)
\label{fig:GodXMol}}
\end{figure}

Having established the promising nature of 18c6-nap spectroscopically, we proceeded to single molecule-level tests. Using a similar technique to Ref~\cite{McDonald:2017izm}, a poly-vinyl alcohol (PVA) gel matrix suspending barium enriched 18c6-nap was imaged using total internal reflection fluorescence (TIRF) imaging.  A single barium ion image taken with our new dry-compatible molecule is shown in Fig~\ref{fig:GodXMol}, left, with the single-step photobleaching process that is the hallmark of single molecule detection shown in Fig.~\ref{fig:GodXMol}, right. In this environment, 18c6-nap shows a considerably brighter response to single ions than the originally investigated dyes of Ref~\cite{Jones:2016qiq,McDonald:2017izm}, with each spot showing an enhanced brightness by a factor of around 12 relative to backgrounds in previous work. 

Notably, the PVA matrix still represents an aqueous environment.  Ongoing work focuses on advancing to a technique for observing single molecules in a true dry state, as described in the next section.  Furthermore, we find that 18c6-nap as a barium receptor has such a strong barium affinity that any naturally occurring barium backgrounds can often be bound to the dye during production, and are difficult to remove. While this behavior is ideal for the efficient capture of barium in xenon, the affinity between the ion and the dye is strong enough that commercially available non-fluorescent chelators such as BAPTA, EDTA and EGTA are unable to outcompete the dye for barium, unlike with previously explored fluorophores which could be ``cleaned'' with those compounds. As such, we are exploring both stronger non-fluorescent chelators and also weaker fluorophore binding groups, in order to out-compete the active fluorophore and realize a darker initial-state.

\begin{figure}
\begin{centering}

\includegraphics[width=0.99\columnwidth]{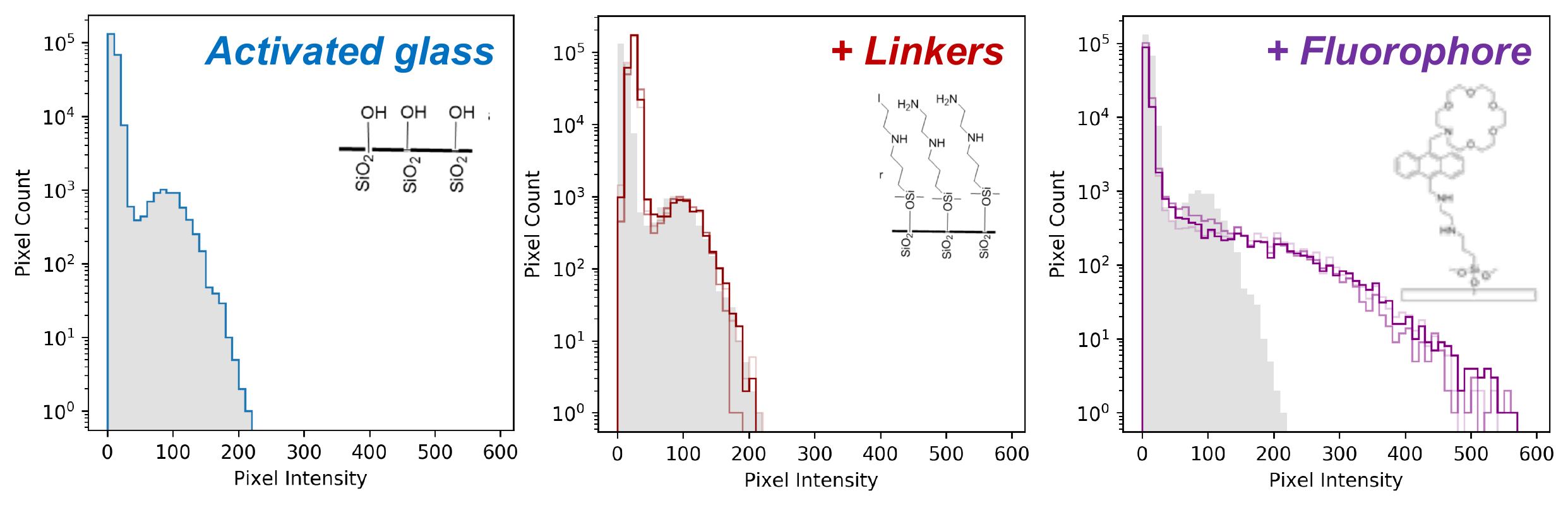}

\par\end{centering}
\caption{Progression of the number of EMCCD camera pixels capturing at specific intensities from cleaned activated glass (left), glass with the hydrocarbon linkers (center), and glass with 18c6-an attached, and activated by ions (right).
\label{fig:Monol}}
\end{figure}

\section{Construction of Self Assembled Monolayers}

As described in our previous works~\cite{Jones:2016qiq, Thapa:2019zjk, Byrnes:2019jxr}, our intention is to generate a self-assembled monolayer of barium responsive fluorescent molecules, rather than a set embedded in a thick matrix, as in recent studies. Adding surface-tethering groups for surface monolayer formation can interfere with fluorescence and binding, and thus for the single molecule studies described in the previous section we used molecules with fully formed tether groups attached to the receptor.  Strong single molecule response was obtained with these fully functionalized molecules, representing an important step toward sensor development.

By silanizing the surface of our glass coverslip and using hydrocarbon linkers to connect an unchelated 18c6 dye to the surface, we then evenly distribute the molecules into a uniform monolayer.  In initial studies using 18c6-anthracene, clean glass, glass with the linkers attached, and glass with the linkers and chelated dye were studied using TIRF microscopy. The clean glass and silanized glass with linkers were shown to have very similar intensity profiles, well within experimental error of one another, whereas the glass with the dye attached and chelated to ions shows a much stronger intensity profile, as is shown in Fig 4. This work is now being advanced with 18c6-nap, with first results to appear in a forthcoming publication.

\section{High Pressure Gas Sensor Development}
In addition to the development of suitable chemistry techniques, concurrent efforts are required to implement tools to extract and analyze the barium ions in high pressure xenon gas. There are two primary challenges-- creation of a method to transport the ionized barium to the image scanning plane, and development of an imaging system that functions in high pressure environments at the single molecule level. 

\subsection{RF Carpet Development}

The first challenge, lateral transport of the barium ions as they near the cathode, is being approached with high-pressure RF carpets~\cite{schwarz2011rf} (these have some similarities to the ``RF Funnels'' explored by the nEXO collaboration~\cite{Thapa:2019zjk}, but employing in-situ detection rather than gas flow and extraction to vacuum). Barium ions produced in bulk xenon gas are to be drifted to the cathode via a DC field~\cite{Bainglass:2018odn}. Here, they will be levitated by RF waves and swept inwards towards the center where an SMFI scanning plane will be situated, with dimensions of a few mm$^2$. R\&D on RF carpet implementation in high pressure gases is ongoing.

\begin{figure}
\begin{centering}
\includegraphics[width=0.55\columnwidth]{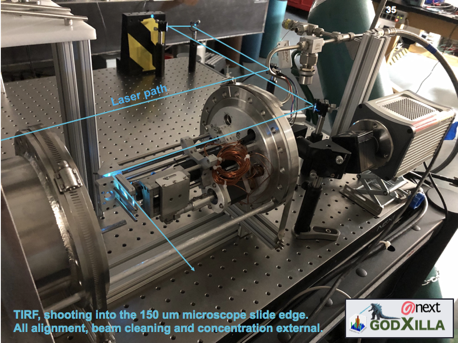}
\includegraphics[width=0.43\columnwidth]{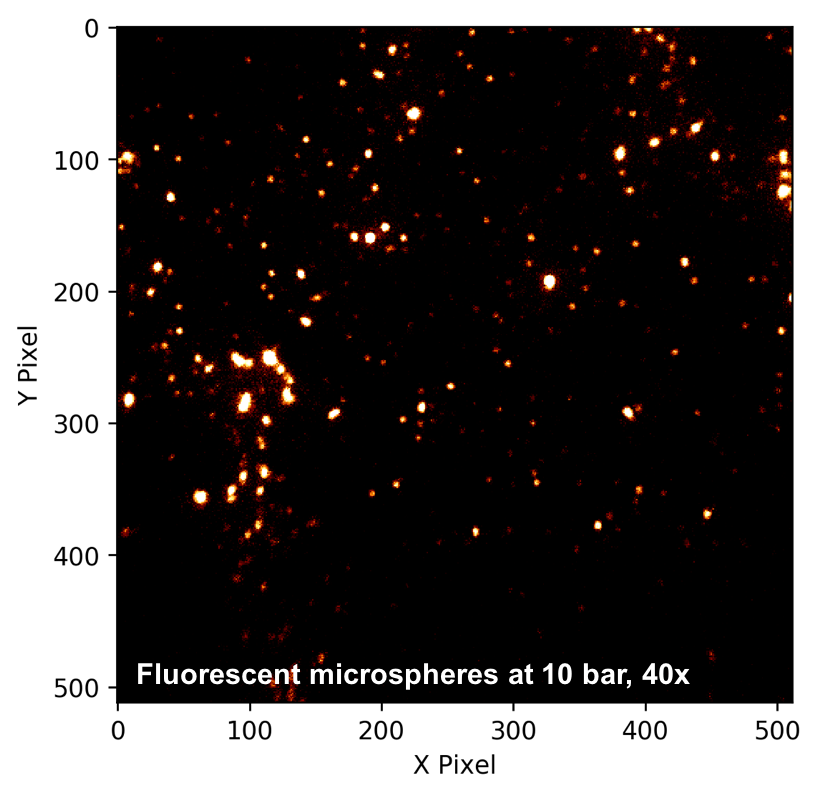}
\par\end{centering}
\caption{Prototype high-pressure gas microscope at the University of Texas Arlington, laser  path in blue (left) and an image taken using this prototype microscope with a 40x air coupled objective showing fluorescent green-yellow microspheres at 10 bar (right).
\label{fig:Microscope}}
\end{figure}

\subsection{High Pressure Microscope Development}

The second technological challenge is the design of a high-pressure compatible microscope capable of detecting a single molecules. Due to the short working distance of high magnification objectives, both the scanning surface and objective must be present within the chamber. All electronics, including the electron multiplying CCD camera and the excitation laser can be housed externally, minimizing the number of elements that have to be made pressure compatible. A prototype of this concept has been constructed and tested, demonstrating high resolution fluorescence microscopy at pressures up to ten bar using 20x and 40x Olympus air coupled objectives, custom-modified in house to be robust against high pressures.  A photograph of this setup is shown in Fig.~\ref{fig:Microscope}, left, as well as a calibration image of fluorescent microspheres in Fig.~\ref{fig:Microscope}, right, showing excellent resolution at 10 bar and with fully external optics. The focusing of the objective is controlled by a stepper motor capable of micrometer incremental movement, and all optical paths enter and leave the vessel using sapphire windows. The next step in the development of this novel microscope apparatus will be to image individual molecules in high pressure gas, realization of which is ongoing.

\section{Barium Tagging R\&D in NEXT}

We have demonstrated dry-capable, single molecule resolving, barium selective fluorophores for daughter ion imaging in high pressure xenon gas.  We have also assembled a high pressure gas microscope system which is under optimization for single ion resolution. These components, in concert with a small RF carpet, comprise the ``NEXT-GodXilla'' program at the University of Texas at Arlington.  The goal of NEXT-GodXilla is to demonstrate identification of single barium ion candidates impinging on SMFI sensors in high pressure xenon gas.

The results described in this paper are part of the NEXT R\&D program on Barium Tagging which comprises activities in the USA, Spain and in Israel.  In Spain, the approach to dry sensors is being pursued through the development of bicolor indicators, and two photon microscopy is being investigated as sensor scanning technique \cite{DIPCMolecule}. In Israel, and also in collaboration with CERN-Isolde a decelerated Ba$^{2+}$ ion beam is under development. Systems designed to prove the transport and detection of Ba$^{2+}$ in gas are under development at UTA (GodXilla) and at the Canfranc Underground Laboratory (RITA). A program to develop RF carpets for concentration of ions to sensors in high pressure gases is ongoing, in collaboration with Argonne National Laboratory, where selected and decelerated barium ions from the CARIBU rare isotope breeder will be used to test ion delivery to SMFI sensors.  In parallel, efforts to develop a movable ion capture system are underway in Europe and Israel. 

In the medium-term, the NEXT collaboration plans to employ a single barium ion sensor within the existing NEXT-White~\cite{Monrabal:2018xlr} detector, once the NEXT-100 experiment~\cite{Martin-Albo:2015rhw} (presently in construction) begins taking data.  To achieve this goal, R\&D continues in order to overcome the remaining challenges involved in realizing barium tagging in high pressure xenon gas detectors.

\section*{Acknowledgements}
This work was supported by the Department of Energy under Early Career Award number DE-SC0019054. The University of Texas at Arlington NEXT group is also supported by Department of Energy Award DE-SC0019223.

\bibliography{main}

\begin{thebibliography}{10}
\expandafter\ifx\csname url\endcsname\relax
  \def\url#1{\texttt{#1}}\fi
\expandafter\ifx\csname urlprefix\endcsname\relax\def\urlprefix{URL }\fi
\expandafter\ifx\csname href\endcsname\relax
  \def\href#1#2{#2} \def\path#1{#1}\fi

\bibitem{Schechter:1980gr}
J.~Schechter, J.~W.~F. Valle, {Neutrino Masses in SU(2) x U(1) Theories}, Phys.
  Rev. D22 (1980) 2227.
\newblock \href {http://dx.doi.org/10.1103/PhysRevD.22.2227}
  {\path{doi:10.1103/PhysRevD.22.2227}}.

\bibitem{Vergados:2012xy}
J.~D. Vergados, H.~Ejiri, F.~Simkovic, {Theory of Neutrinoless Double Beta
  Decay}, Rept. Prog. Phys. 75 (2012) 106301.
\newblock \href {http://arxiv.org/abs/1205.0649} {\path{arXiv:1205.0649}},
  \href {http://dx.doi.org/10.1088/0034-4885/75/10/106301}
  {\path{doi:10.1088/0034-4885/75/10/106301}}.

\bibitem{Bilenky:2012qi}
S.~M. Bilenky, C.~Giunti, {Neutrinoless double-beta decay: A brief review},
  Mod. Phys. Lett. A27 (2012) 1230015.
\newblock \href {http://arxiv.org/abs/1203.5250} {\path{arXiv:1203.5250}},
  \href {http://dx.doi.org/10.1142/S0217732312300157}
  {\path{doi:10.1142/S0217732312300157}}.

\bibitem{Chang:1985en}
D.~Chang, R.~N. Mohapatra, {Comment on the 'Seesaw' Mechanism for Small
  Neutrino Masses}, Physical Review D32 (1985) 1248.
\newblock \href {http://dx.doi.org/10.1103/PhysRevD.32.1248}
  {\path{doi:10.1103/PhysRevD.32.1248}}.

\bibitem{Fukugita:1986hr}
M.~Fukugita, T.~Yanagida, {Baryogenesis Without Grand Unification}, Phys. Lett.
  B174 (1986) 45--47.
\newblock \href {http://dx.doi.org/10.1016/0370-2693(86)91126-3}
  {\path{doi:10.1016/0370-2693(86)91126-3}}.

\bibitem{KamLAND-Zen:2016pfg}
A.~Gando, et~al., {Search for Majorana Neutrinos near the Inverted Mass
  Hierarchy Region with KamLAND-Zen}, Phys. Rev. Lett. 117~(8) (2016) 082503,
  [Addendum: Phys. Rev. Lett.117,no.10,109903(2016)].
\newblock \href {http://arxiv.org/abs/1605.02889} {\path{arXiv:1605.02889}},
  \href {http://dx.doi.org/10.1103/PhysRevLett.117.109903,
  10.1103/PhysRevLett.117.082503} {\path{doi:10.1103/PhysRevLett.117.109903,
  10.1103/PhysRevLett.117.082503}}.

\bibitem{Moe:1991ik}
M.~K. Moe, {New approach to the detection of neutrinoless double beta decay},
  Physical Review C44 (1991) 931--934.
\newblock \href {http://dx.doi.org/10.1103/PhysRevC.44.931}
  {\path{doi:10.1103/PhysRevC.44.931}}.

\bibitem{Renner:2019pfe}
J.~Renner, et~al., {Energy Calibration of the NEXT-White Detector with 1\%
  Resolution Near Q$_{\beta\beta}$ of $^{136}$Xe }, Accepted by JHEP\href
  {http://arxiv.org/abs/1905.13110} {\path{arXiv:1905.13110}}.

\bibitem{Ferrario:2019kwg}
P.~Ferrario, et~al., {Demonstration of the Event Identification Capabilities of
  the NEXT-White Detector}, Accepted by JHEP\href
  {http://arxiv.org/abs/1905.13141} {\path{arXiv:1905.13141}}.

\bibitem{Nygren:2009zz}
D.~Nygren, {High-pressure xenon gas electroluminescent TPC for 0--$\nu$ $\beta
  \beta$ --decay search}, Nucl. Instrum. Meth. A603 (2009) 337--348.
\newblock \href {http://dx.doi.org/10.1016/j.nima.2009.01.222}
  {\path{doi:10.1016/j.nima.2009.01.222}}.

\bibitem{Martin-Albo:2015rhw}
J.~Mart{\'\i}n-Albo, et~al., {Sensitivity of NEXT-100 to Neutrinoless Double
  Beta Decay}, JHEP 05 (2016) 159.
\newblock \href {http://arxiv.org/abs/1511.09246} {\path{arXiv:1511.09246}},
  \href {http://dx.doi.org/10.1007/JHEP05(2016)159}
  {\path{doi:10.1007/JHEP05(2016)159}}.

\bibitem{stuurman2006imaging}
N.~Stuurman, R.~Vale, Imaging single molecules using total internal reflection
  fluorescence microscopy (2006).

\bibitem{Elba}
D.~Nygren, Detection of the barium daughter in $^{136}$\uppercase{X}e to
  $^{136}$\uppercase{B}a + 2e- by in situ single-molecule fluorescence imaging,
  Nuclear Instruments and Methods in Physics Research Section A: Accelerators,
  Spectrometers, Detectors and Associated Equipment 824 (2016) 2--5.

\bibitem{Jones:2016qiq}
B.~Jones, A.~McDonald, D.~Nygren, Single molecule fluorescence imaging as a
  technique for barium tagging in neutrinoless double beta decay, Journal of
  Instrumentation 11~(12) (2016) P12011.

\bibitem{McDonald:2017izm}
A.~D. McDonald, et~al., {Demonstration of Single Barium Ion Sensitivity for
  Neutrinoless Double Beta Decay using Single Molecule Fluorescence Imaging},
  Phys. Rev. Lett. 120~(13) (2018) 132504.
\newblock \href {http://arxiv.org/abs/1711.04782} {\path{arXiv:1711.04782}},
  \href {http://dx.doi.org/10.1103/PhysRevLett.120.132504}
  {\path{doi:10.1103/PhysRevLett.120.132504}}.

\bibitem{Thapa:2019zjk}
P.~Thapa, I.~Arnquist, N.~Byrnes, A.~A. Denisenko, F.~W. Foss, B.~J.~P. Jones,
  A.~D. Mcdonald, D.~R. Nygren, K.~Woodruff, {Barium Chemosensors with
  Dry-Phase Fluorescence for Neutrinoless Double Beta Decay}, Accepted by
  Nature Sci. Rep.\href {http://arxiv.org/abs/1904.05901}
  {\path{arXiv:1904.05901}}.

\bibitem{Byrnes:2019jxr}
N.~Byrnes, F.~W. Foss, B.~J.~P. Jones, A.~D. McDonald, D.~R. Nygren, P.~Thapa,
  A.~Trinidad, {Progress toward Barium Tagging in High Pressure Xenon Gas with
  Single Molecule Fluorescence Imaging}, in: {9th Symposium on Large TPCs for
  Low-Energy Rare Event Detection (TPC2018) Paris, France, December 12-14,
  2018}, 2019.
\newblock \href {http://arxiv.org/abs/1901.03369} {\path{arXiv:1901.03369}}.

\bibitem{schwarz2011rf}
S.~Schwarz, Rf ion carpets: The electric field, the effective potential,
  operational parameters and an analysis of stability, International Journal of
  Mass Spectrometry 299~(2-3) (2011) 71--77.

\bibitem{Bainglass:2018odn}
E.~Bainglass, B.~J.~P. Jones, F.~W. Foss, M.~N. Huda, D.~R. Nygren, {Mobility
  and Clustering of Barium Ions and Dications in High Pressure Xenon Gas},
  Phys. Rev. A97~(6) (2018) 062509.
\newblock \href {http://arxiv.org/abs/1804.01169} {\path{arXiv:1804.01169}},
  \href {http://dx.doi.org/10.1103/PhysRevA.97.062509}
  {\path{doi:10.1103/PhysRevA.97.062509}}.

\bibitem{DIPCMolecule}
I.~Rivilla, et~al., {Towards a background-free neutrinoless double beta decay
  experiment based on a fluorescent bicolor sensor, }\href
  {http://arxiv.org/abs/1909.02782} {\path{arXiv:1909.02782}}.

\bibitem{Monrabal:2018xlr}
F.~Monrabal, et~al., {The Next White (NEW) Detector}, JINST 13~(12) (2018)
  P12010.
\newblock \href {http://arxiv.org/abs/1804.02409} {\path{arXiv:1804.02409}},
  \href {http://dx.doi.org/10.1088/1748-0221/13/12/P12010}
  {\path{doi:10.1088/1748-0221/13/12/P12010}}.

\end{thebibliography}
\bibliographystyle{elsarticle-num}

\end{document}